\documentclass[aps,twocolumn,prb,superscriptaddress,amsmath,amssymb]{revtex4-1}
\usepackage{graphicx}
\usepackage{dcolumn}
\usepackage{bm}
\usepackage{float}
\usepackage{placeins}
\usepackage{hyperref}
\usepackage{hhline}
\usepackage{multirow}
\usepackage{array}
\usepackage{tabularx}
\usepackage{booktabs}
\usepackage{romannum}
\usepackage{amsmath}
\usepackage{dsfont} 
\usepackage{ifthen}
\usepackage[dvipsnames]{xcolor}
\usepackage{soul}
\usepackage{ulem} 
\hypersetup{colorlinks=true,linkcolor=blue,filecolor=magenta,urlcolor=blue,citecolor=blue,pdftitle={Overleaf Example}}

\renewcommand{\figurename}{\textbf{Fig.}}

\renewcommand{\tablename}{\textbf{Table}}

\begin{document}

\title{Intertwined Orders in a Quantum-Entangled Metal}

\author{Junyoung~Kwon}
\altaffiliation{These authors contributed equally to this work.}
\affiliation{Department of Physics, Pohang University of Science and Technology, Pohang 37673, Korea}

\author{Jaehwon~Kim}
\altaffiliation{These authors contributed equally to this work.}
\affiliation{Department of Physics, Pohang University of Science and Technology, Pohang 37673, Korea}

\author{Gwansuk Oh}
\altaffiliation{These authors contributed equally to this work.}
\affiliation{Department of Physics, Pohang University of Science and Technology, Pohang 37673, Korea}

\author{Seyoung~Jin}
\altaffiliation{These authors contributed equally to this work.}
\affiliation{Department of Physics, Pohang University of Science and Technology, Pohang 37673, Korea}
\affiliation{Center for Artificial Low Dimensional Electronic Systems, Institute for Basic Science, Pohang, 37673, Korea}

\author{Kwangrae~Kim}
\altaffiliation{These authors contributed equally to this work.}
\affiliation{Department of Physics, Pohang University of Science and Technology, Pohang 37673, Korea}

\author{Hoon~Kim}
\altaffiliation{These authors contributed equally to this work.}
\affiliation{Department of Physics, Pohang University of Science and Technology, Pohang 37673, Korea}

\author{Seunghyeok~Ha}
\affiliation{Department of Physics, Pohang University of Science and Technology, Pohang 37673, Korea}

\author{Hyun-Woo J.~Kim}
\affiliation{Department of Physics, Pohang University of Science and Technology, Pohang 37673, Korea}

\author{GiBaik Sim}
\affiliation{Department of Physics, Hanyang University, Seoul 04763, Korea}

\author{Bj{\"o}rn Wehinger}
\affiliation{ESRF, The European Synchrotron, 71 Avenue des Martyrs, CS40220, 38043 Grenoble Cedex 9, France}
\author{Gaston Garbarino}
\affiliation{ESRF, The European Synchrotron, 71 Avenue des Martyrs, CS40220, 38043 Grenoble Cedex 9, France}

\author{Nour Maraytta}
\affiliation{Institute for Quantum Materials and Technologies, Karlsruhe Institute of Technology, Karlsruhe 76021,Germany}

\author{Michael Merz}
\affiliation{Institute for Quantum Materials and Technologies, Karlsruhe Institute of Technology, Karlsruhe 76021,Germany}
\affiliation{Karlsruhe Nano Micro Facility, Karlsruhe Institute of Technology, Eggenstein-Leopoldshafen 76344,Germany}

\author{Matthieu Le Tacon}
\affiliation{Institute for Quantum Materials and Technologies, Karlsruhe Institute of Technology, Karlsruhe 76021,Germany}

\author{Christoph~J.~Sahle}
\affiliation{ESRF, The European Synchrotron, 71 Avenue des Martyrs, CS40220, 38043 Grenoble Cedex 9, France}
\author{Alessandro~Longo}
\affiliation{ESRF, The European Synchrotron, 71 Avenue des Martyrs, CS40220, 38043 Grenoble Cedex 9, France}
\affiliation{Istituto per lo Studio dei Materiali Nanostrutturati - Consiglio Nazionale delle Ricerche, Palermo 90146, Italy}

\author{Jungho Kim}
\affiliation{Advanced Photon Source, Argonne National Laboratory, Lemont, IL 60439, USA}

\author{Ara Go}
\affiliation{Department of Physics, Chonnam National University, Gwangju 61186, Korea}

\author{Gil~Young~Cho}
\email[email: ]{gilyoungcho@kaist.ac.kr}
\affiliation{Department of Physics, Korea Advanced Institute of Science and Technology, Daejeon 34141, Korea}
\affiliation{Center for Artificial Low Dimensional Electronic Systems, Institute for Basic Science, Pohang, 37673, Korea}
\affiliation{Asia-Pacific Center for Theoretical Physics, Pohang, 37673, Korea}

\author{Beom~Hyun~Kim}
\email[email: ]{bomisu@gmail.com}
\affiliation{Center for Theoretical Physics of Complex Systems, Institute for Basic Science, Daejeon 34126, Korea}
\affiliation{Department of Physics and Astronomy, Seoul National University, Seoul 08826, Korea}%

\author{B.~J. Kim}
\email[email: ]{bjkim6@postech.ac.kr}
\affiliation{Department of Physics, Pohang University of Science and Technology, Pohang 37673, Korea}

\date{\today}

\maketitle

\noindent
{\bf Entanglement underpins quantum information processing and computing~\cite{Hor09}, yet its experimental quantification in complex, many-body condensed matter systems remains a considerable challenge~\cite{Ami08,Lau24}. Here, we reveal a highly entangled electronic phase proximate to a quantum metal-insulator transition, identified by resonant inelastic x-ray scattering interferometry~\cite{Rev19}. This approach reveals that entanglement across atomic sites generates characteristic interference patterns, which our model accurately reproduces, enabling extraction of a full entanglement spectrum and resolution of the underlying quantum states. Our analysis of the pyrochlore iridate Nd$_2$Ir$_2$O$_7$ demonstrates that the system undergoes pronounced quantum fluctuations in its spin, orbital and charge degrees of freedom, even in the presence of a long-range `all-in-all-out' antiferromagnetic order~\cite{Tom12,Guo16}. Importantly, the observed entanglement signatures facilitate the coexistence of multiple exotic symmetry‐breaking orders. Complementary investigations using Raman spectroscopy corroborate the presence of these hidden orders and their emergent excitations. In particular, we observe a two-magnon-bound state~\cite{Bet31, Han63, Wor63, Tor69} below the lowest single-magnon excitation energy, which, together with split phonon modes, provides strong evidence for cubic symmetry-breaking orders of magnetic origin~\cite{Chu91,Sha06, Zhi10,Pra20, Fog24, She25} juxtaposed with the all-in-all-out order. Our work thus establishes a direct link between quantum entanglement and emergent unconventional orders, opening new avenues for investigating quantum materials.
}

\noindent
Entanglement is a fundamental phenomenon in quantum-mechanical systems, where pairs or groups of particles interact in such a way that the quantum state of each particle cannot be described independently of the states of the others, even when the particles are separated by large distances. Although counterintuitive, this aspect of quantum theory is empirically verified~\cite{Asp82} and has significant implications for technology and our understanding of the fundamental principles of the universe. In particular, entanglement is crucial for understanding quantum phase transitions, which occur at absolute zero temperature. These transitions, driven by quantum fluctuations, involve changes in the ground state of a system~\cite{Sac99}, such as from a magnetically ordered phase to a quantum-disordered phase~\cite{Chu94}. The properties of entanglement provide a detailed picture of these transformations, reflecting the fundamental shifts in the quantum state of the system. Although full experimental access to the structure of entanglement remains challenging in condensed matter systems, measures of entanglement, such as one-tangle, two-tangle and quantum Fisher information, have been extracted from correlation functions measured by spectroscopic techniques~\cite{Sch21,Hal23,Ren24}. While they give valuable insight into the nature and structure of the entanglement, these witnesses provide only a single number to indicate whether the entanglement is present or give a measure of its strength. 

\begin{figure*}
\centering
\includegraphics[width=1.85\columnwidth]{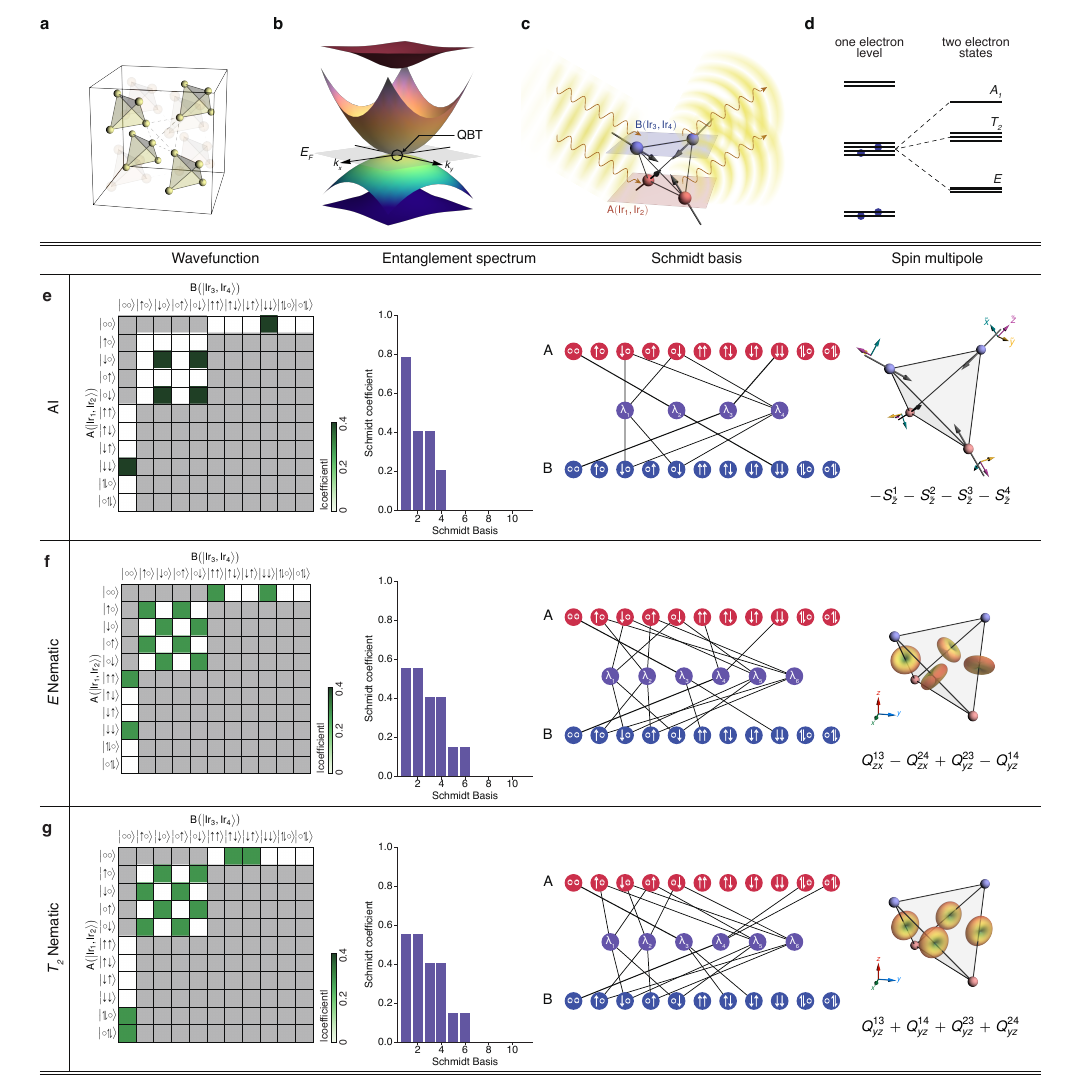}

\caption{\label{fig:wide} {\bf Possible emergent orders and their entanglement structures. a,} Crystal structure of Nd$_2$Ir$_2$O$_7$ in which the pyrochlore network formed by Ir ions (yellow sphere) interpenetrates that of Nd ions (light orange). {\bf b,} Electronic band structure featuring quadratic band touching (QBT) at the Fermi level.  {\bf c,} Bipartition of the tetrahedron into subsystems A(red) and B(blue). {\bf d,} Electronic levels of an isolated tetrahedron. {\bf e-h,} Wavefunction, Schmidt coefficient, Schmidt basis, and spin multipoles for states with AI ({\bf e}), $E$ nematic ({\bf f}), and $T_2$ nematic ({\bf g}) orders. Local coordinate axes, $\tilde{x}$ (green), $\tilde{y}$ (yellow), and $\tilde{z}$ (red), for each Ir site are  shown in {\bf e}. Global coordinate axes $x$, $y$, and $z$ are shown in {\bf f} and {\bf g}. Gray area in the wavefunctions shown in ({\bf e}-{\bf g}) denote unphysical states.
}
\end{figure*}

In this Article, we demonstrate a novel method based on interferometry that captures the full complexity of quantum entanglement and enables the prediction of new emergent orders. The pyrochlore iridates, whose lattice and electronic band structures are shown in Figs.~1a and ~1b, respectively, are an ideal platform to investigate the entanglement structure for several reasons. First, the system can be tuned in proximity to a quantum metal-insulator transition through the radius of the rare-earth ion~\cite{Yan01,Ued16}; among pyrochlore iridates Nd$_2$Ir$_2$O$_7$ exhibits the lowest transition temperature ($T_\textrm{C}$$\approx 33$ K), which delineates a high-temperature metallic phase from a low-temperature insulating phase hosting the `all-in-all-out' (AIAO) magnetic order (Fig.~1e)~\cite{Tom12,Sag13,Dis14,Guo16}. Second, earlier theoretical studies~\cite{Abr74,Wan11_DFT,Wit12,Moo13,Sav14,DasSarma2017, Herbut2017, Moessner2021} have shown that Coulomb interactions become significant due to the peculiar band structure with quadratic band touching (QBT) at the Fermi level (Fig.~1b)~\cite{Nak16}, thereby enabling quantum fluctuations among various distinct phases. This study reveals an additional hidden order that coexists with the AIAO order, as discussed below. Third, its structural motif, a tetrahedron with iridium ions at the vertices, allows for direct experimental access to entanglement using resonant inelastic x-ray scattering (RIXS). We bipartition the tetrahedron as shown in Fig.~1c and measure the entanglement between the two subsystems labeled A and B.

Before discussing our data, we first illustrate, using a simple model on a tetrahedron, how a state with a rich entanglement structure supporting multiple symmetry-breaking orders can arise when spin and charge fluctuations become significant. The pyrochlore lattice is generated by placing a tetrahedron on every lattice point of a face-centered cubic Bravais lattice. In the Mott insulating limit, the states with fully saturated AO and AI orders are represented by 
\begin{align}
    |\uparrow,\uparrow\rangle_A \otimes |\uparrow,\uparrow\rangle_B \quad
 \textrm{and}\quad |\downarrow,\downarrow\rangle_A \otimes |\downarrow,\downarrow\rangle_B,
\end{align}
respectively, where $\uparrow$($\downarrow$) denotes an up-spin(down-spin) state along the local $z$-axis, as defined in Fig.~1e, for each electron localized at four Ir sites. 
These states, being simple direct product states, clearly possess no entanglement. By contrast, weak- or intermediate-coupling descriptions necessitate entangled states to express the AIAO order. To see this point, it is instructive to consider the non-interacting limit. In this picture, the electrons fill the `molecular' orbitals of the tetrahedron in such a way that only two electrons per tetrahedron contribute to magnetism (Fig.~1d). The 2-4-2 orbital degeneracy sequence corresponds to the band degeneracies at the $\Gamma$ point. Written in the same basis as that used in Eq.~(1), each Ir site can be in one of the four states: empty($\circ$), singly occupied ($\uparrow$ or $\downarrow$), or doubly occupied ($\uparrow\downarrow$). 
The two-electron wavefunctions supporting the maximal AIAO orders, which have the same symmetry as in the Mott insulating limit (transforming as $E$ doublet) are

\begin{equation}
\begin{aligned}
|\Psi\rangle_\textrm{AI} &=
 +0.7887\frac{1}{\sqrt{2}} \bigl(|\downarrow,\circ\rangle + |\circ,\downarrow\rangle\bigr)_A 
   \otimes \frac{1}{\sqrt{2}} \bigl(|\downarrow,\circ\rangle - |\circ,\downarrow\rangle\bigr)_B \\
&\quad + 0.4082\,|\circ, \circ\rangle_A \otimes |\downarrow,\downarrow\rangle_B \\
&\quad - 0.4082\,i\,|\downarrow,\downarrow\rangle_A \otimes |\circ, \circ\rangle_B \\
&\quad + 0.2113\frac{i}{\sqrt{2}} \bigl(|\downarrow,\circ\rangle - |\circ,\downarrow\rangle\bigr)_A 
   \otimes \frac{1}{\sqrt{2}} \bigl(|\downarrow,\circ\rangle + |\circ,\downarrow\rangle\bigr)_B 
\end{aligned}
\end{equation}
and its time-reversal pair (Supplementary Note 1). These states are expressed in a basis that requires a minimal number of product states, which is obtained through the process known as Schmidt decomposition (Supplementary Note 2). The number of required basis states, or the Schmidt number, quantifies the degree of entanglement. The wavefunction is visualized in Fig.~1e, along with the absolute value of the Schmidt coefficients and the quantum states comprising each Schmidt basis state. 

Importantly, within this six-dimensional Hilbert space (spanning six distinct ways of occupying the four molecular orbitals with two electrons) states supporting additional orders can be constructed. For instance, a state that breaks the cubic symmetry of the pyrochlore lattice can be derived from the same basis states that generate AI and AO states. This state, shown in Fig.~1f, exhibits quadrupoles $Q_{zx}^{13}-Q_{zx}^{24}+Q_{yz}^{23}-Q_{yz}^{14}$ on the bonds of $E$ symmetry, along with various other order parameters of the same symmetry, including vector chiral and valence bond orders (Supplmentary Note 3). Thus, the entanglement structure within these basis states dictates the symmetry of the resulting order parameters. Similarly, a $T_2$ symmetry state can also be constructed (Fig.~1g). A complete classification of possible orders is provided in the Supplementary Note 3. These orders can coexist with AIAO order as shown in Extended Data Fig.~1.

\vspace{5 mm}
\noindent
\textbf{RIXS interferometry}

Bipartite entanglement manifests itself as interference of inelastically scattered x-rays from subsystems A and B (Fig.~1c). This interference arises only when two or more scattering processes share the same initial and final states. It follows that at least one of these states must be entangled to produce an interference pattern. If, instead, both the initial and the final states were separable (i.e., not entangled), then scattering from subsystem A(B) would leave the state of B(A) unchanged. Consequently, the two processes would lead to distinct final states and would not interfere, except in the special case where the final state is identical to the initial state; i.e., elastic scattering (Supplementary Note 4). The interference produces a sinusoidal intensity oscillation as a function of x-ray path length difference, or equivalently, the momentum transfer along the $L$ reciprocal lattice vector in our setup (Methods).

\begin{figure*}
\centering
\includegraphics[width=1.85\columnwidth]{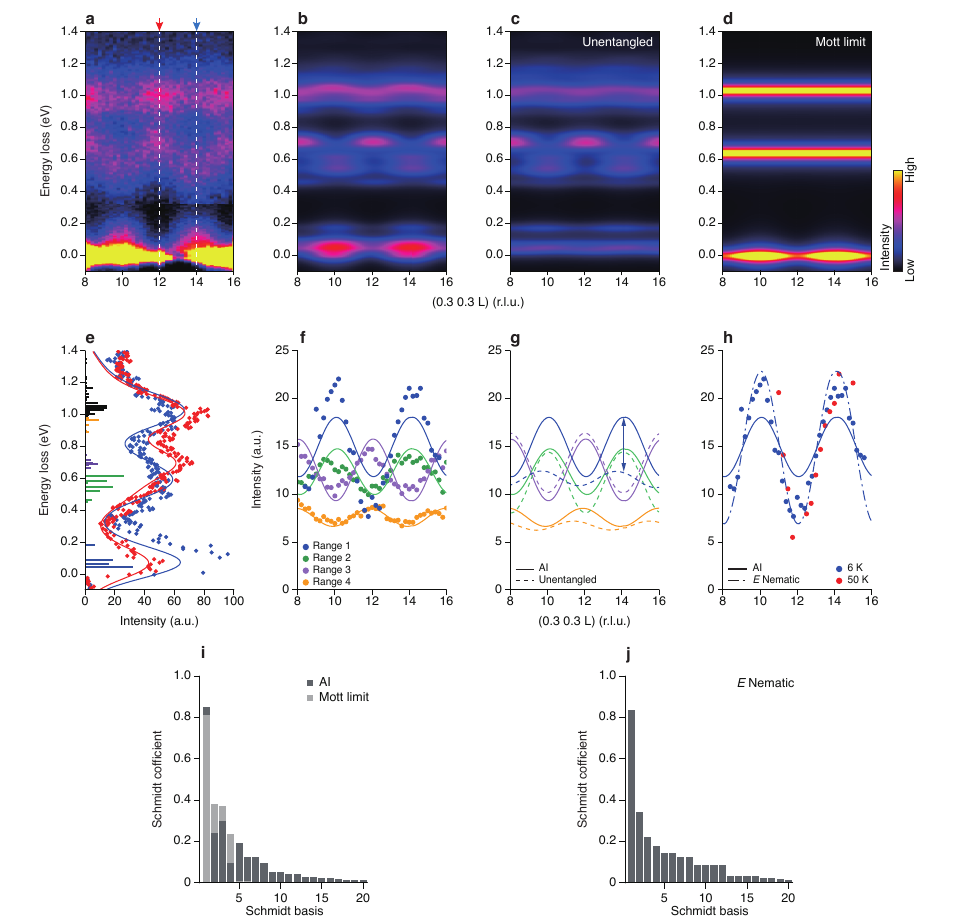}
\caption{\label{fig:wide} {\bf Extraction of the ground-state entanglement. a,} RIXS spectra measured at $T=$6 $K$. Energy spectra are shown in {\bf e} for the two momenta indicated by dotted vertical lines. {\bf b,} Calculation of RIXS spectra using exact diagonalization on a cluster with parameters that best fit the experimental data. {\bf c,} Calculation for an unentangled ground state. {\bf d,} Calculation for the state approaching the Mott limit. {\bf e,} Comparison of measured (dots) and calculated energy spectra (solid lines) at $L$=12 (red) and $L$=14 (blue). The energies and intensities of the calculated final states before Gaussian convolution are shown using  sticks, and their colors indicate four different energy windows for integration. {\bf f-h} Integrated intensities (dots) as a function of $L$ compared to calculations (solid and dashed lines) for the ground states with AIAO order ({\bf f}), no entanglement ({\bf g}), and $E$ nematic order ({\bf h}). In {\bf h}, the data measured at T=50 K (red) is also shown. {\bf i,j,} Entanglement spectra for the ground state with AI ({\bf i}) and $E$ nematic ({\bf j}) orders.
}
\end{figure*}

Figure 2a presents the measured RIXS intensity at the base temperature of $T=$6 $K$ over a wide range of $L$, revealing clear oscillations in intensity, with amplitudes and phases dependent on the final states. The period is four reciprocal lattice units in $L$, consistent with the fact that the separation between subsystems A and B is one-quarter of the lattice parameter ($a$$\sim$10.3815 {\AA}) of the conventional face-centered-cubic unit cell. Inspection of the energy spectra shown for $L$=12 and $L$=14 (Fig.~2e), clearly shows that the phase becomes $\pi$-shifted at around 0.6 eV. As shown in Fig.~2b, this feature is well captured by our model calculations using exact diagonalization on a cluster of a single tetrahedron containing 10,067 basis states, which include all states with energies $\lesssim$ 1 eV (Supplemenatry Note 5). The parameters used in the calculations are listed in Methods. The ground state is an $E$ doublet, consistent with the minimal model in Fig.~1d, and the ground state is initially assumed to be the state that gives the maximal AI order (Extended Data Fig.~2). For a quantitative comparison with the calculations, the measured intensity is integrated over four different energy windows, and the corresponding amplitudes and phases of the oscillations are shown in Fig.~1f. 

To determine which final states exhibit the strongest sensitivity to initial state entanglement, we simulate in Fig.~2c the interference pattern expected if the initial state were not entangled; specifically, we replace it with the single most dominant direct-product AI state. The oscillation amplitude is most strongly suppressed for the lowest-energy states, indicating their high sensitivity to initial-state entanglement, whereas oscillations in the higher-energy states predominantly reflect entanglement in the final states. When the hopping amplitudes are uniformly scaled by a factor $r$=0.05 to bring the system close to the Mott limit (Fig.~2d), the interference pattern disappears for the final states above 0.2 eV, and the Schmidt number decreases to 4 (Fig.~2i). As electrons become localized around individual atoms, entanglement is lost in the high-energy charge excitation states but persists in the low-energy spin sector. This persistance arises because the resulting spin Hamiltonian deviates from the pure Ising limit which yields the ground state in Eq.~(1).

The AI state wavefunction consists of a superposition of many states, with substantial contributions from those involving charge and orbital fluctuations. Intriguingly, the best fit to the experimental data is obtained for the initial state that has $E$ nematic order (Fig.~2h and Extended Data Fig.~2). This state is an equal-weight superposition of the AI and AO states with no dipolar order, in contrast to the AIAO order observed in most of the other members of this material family~\cite{Don16,Sag13}. In fact, we find unambiguous signatures of the AIAO order from resonant x-ray diffraction (Extended Data Fig.~3). This suggests that the interference is not directly sensitive to a specific order parameter but rather to the entanglement structure of the electronic wavefunction associated with the order. Indeed, interference patterns remain unchanged across T$_\textrm{C}$ (Fig.~2h), indicating that the entanglement structure is largely unaffected when the sample is heated above T$_\textrm{C}$. Thus, our data suggest the presence of an additional order characterized by inter-site entanglement, possibly associated with $E$ quadrupolar moments (Fig.~1).

\vspace{5 mm}
\noindent
\textbf{Symmetry considerations} 

The six states in Fig.~1c split into a singlet ($\psi^{A_1}$), a doublet ($\psi^{E}$) and a triplet ($\psi^{T_2}$) when electron interactions are taken into account (Fig.~1d and Supplementary Note 1). The $\psi^{E}$ and $\psi^{T_2}$ states together span the low-energy manifold in which the spins are constrained to align along the local $z$ axes~\cite{Sav14, DasSarma2017, Herbut2017}. The ground state must be $\psi^{E}$, as AIAO order emerges from these states (Supplementary Note 1). The symmetry of possible order parameters $O$ is such that $\langle \psi^{E}|O|\psi^{E}\rangle \neq 0$. Given that $E \otimes E = A_1 \oplus A_2 \oplus E$, two non-trivial orders arise: $A_2$, which transform as an octupole (corresponding to AIAO order), and $E$, which transform as quadrupoles (representing nematic order). However, near a quantum critical point, $\psi^{T_2}$ states can mix into the ground state $\psi^{E}$ doublet, meaning that orders of $T_{1}$ and $T_{2}$ symmetries cannot be ruled out ($T_2 \otimes T_2 = A_1 \oplus E \oplus T_1 \oplus T_2$). Indeed, our mean field calculation shows that both the $E$ and $T_2$ nematic orders can emerge from a QBT model within certain parameter regimes (Fig.~4a and Supplementary Note 6).
The Raman selection rule follows from similar considerations: $\langle \Psi^{f}|R|\Psi^{i}\rangle \neq 0$ when the symmetry of the Raman operator $R$ matches one of the symmetries in the direct product of the symmetries of the initial ($\Psi^{i}$) and the final states ($\Psi^{f}$). Figure 3a and Table 1 summarize the possible excitations in each symmetry channel and the corresponding Raman operators.

\vspace{5 mm}
\noindent
\textbf{Raman spectroscopy}

\begin{figure*}
\centering
\includegraphics[width=1.85\columnwidth]{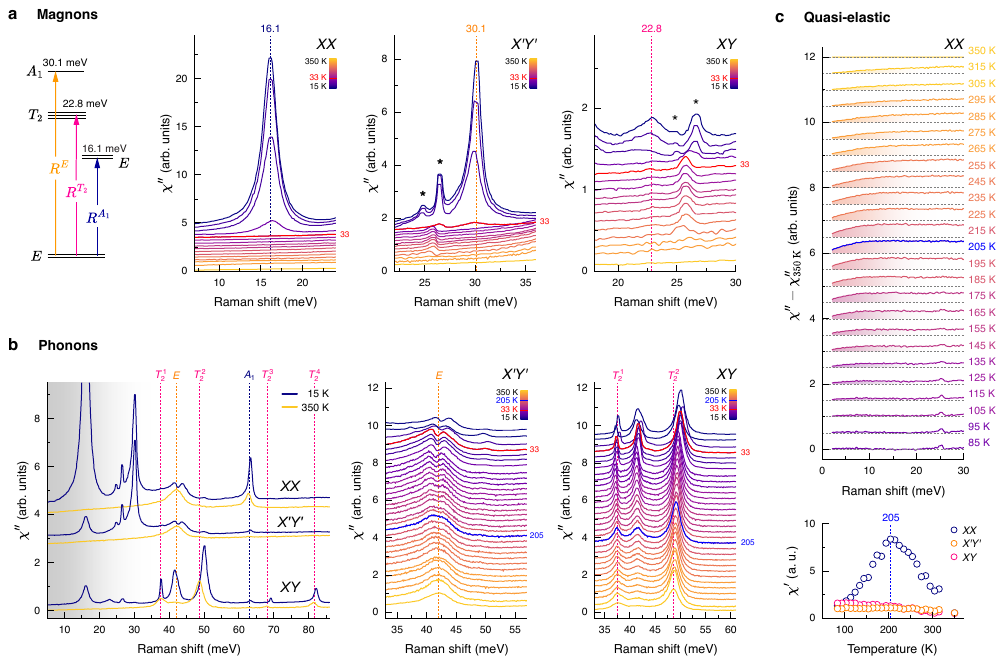}
\caption{\label{fig:wide} {\bf Evidence for cubic symmetry breaking. a,} Schematic of Raman selection rules and spectra measured in XX ($R^{A_1} \oplus R^{E}$), X$'$Y$'$ ($R^{E}$) and XY ($R^{T_2}$) polarization configurations showing magnetic excitations. $R^E$ and $R^{T_2}$ operators are linear combinations of $S_{\Tilde{x}}$ and $S_{\Tilde{y}}$, sensitive to single-magnon excitations, whereas $R^{A_1}$ operator is bilinear in spin operators and involves two spin flips (Supplementary Note 3). Nd CEF excitations are marked by asterisks. {\bf b,} Raman spectra measured in X$'$Y$'$ ($R^{E}$) and XY ($R^{T_2}$) polarization configurations showing phonon excitations (unshaded area).  {\bf c}, Temperature dependence of quasi-elastic electronic scattering (upper), and real part of the susceptibility obtained by Kramers-Kronig transformation of the imaginary part (lower). Error bars are smaller than the size of the markers. 
}
\end{figure*}

\begin{table*}[!hbt]
\begin{ruledtabular}
{\renewcommand{\arraystretch}{1.65}
{\normalsize
\begin{tabular}{cc}
 \rule{1em}{0em}Symmetry\rule{1em}{0em} & \rule{17em}{0em}Raman operators\rule{17em}{0em}  \\
\hline
\hline
 \multirow{5}{*}[.25em]{$R^{A_1}$} & $\mathbf{S}^1\cdot\mathbf{S}^2+\mathbf{S}^1\cdot\mathbf{S}^3+\mathbf{S}^1\cdot\mathbf{S}^4+\mathbf{S}^2\cdot\mathbf{S}^3+\mathbf{S}^2\cdot\mathbf{S}^4+\mathbf{S}^3\cdot\mathbf{S}^4$\\
    & $Q_{3z^2-r^2}^{12}+Q_{3z^2-r^2}^{34}+Q_{3x^2-r^2}^{14}+Q_{3x^2-r^2}^{23}+Q_{3y^2-r^2}^{13}+Q_{3y^2-r^2}^{24}$\\
    & $Q_{xy}^{12}-Q_{xy}^{34}+Q_{zx}^{13}-Q_{zx}^{24}+Q_{yz}^{14}-Q_{yz}^{23}$\\
    & $-(\mathbf{S}^1\times\mathbf{S}^2)_x+(\mathbf{S}^1\times\mathbf{S}^2)_y-(\mathbf{S}^1\times\mathbf{S}^3)_z+(\mathbf{S}^1\times\mathbf{S}^3)_x-(\mathbf{S}^1\times\mathbf{S}^4)_z-(\mathbf{S}^1\times\mathbf{S}^4)_y\hspace{3em}$\\[-.8em]
    & $\hspace{3em}+(\mathbf{S}^2\times\mathbf{S}^3)_y+(\mathbf{S}^2\times\mathbf{S}^3)_z-(\mathbf{S}^2\times\mathbf{S}^4)_x-(\mathbf{S}^2\times\mathbf{S}^4)_z-(\mathbf{S}^3\times\mathbf{S}^4)_x-(\mathbf{S}^3\times\mathbf{S}^4)_y$ \\ 
\hline
\multirow{2}{*}{$R^{E}$} 
    & $S^1_{\Tilde{x}}+S^2_{\Tilde{x}}+S^3_{\Tilde{x}}+S^4_{\Tilde{x}}$\\
    & $S^1_{\Tilde{y}}+S^2_{\Tilde{y}}+S^3_{\Tilde{y}}+S^4_{\Tilde{y}}$\\
\hline
 \multirow{3}{*}{$R^{T_2}$}
    & $S^1_{\Tilde{y}}+S^2_{\Tilde{y}}-S^3_{\Tilde{y}}-S^4_{\Tilde{y}}$\\
    & $(\sqrt{3}S^1_{\Tilde{x}}-S^1_{\Tilde{y}})-(\sqrt{3}S^2_{\Tilde{x}}-S^2_{\Tilde{y}})-(\sqrt{3}S^3_{\Tilde{x}}-S^3_{\Tilde{y}})+(\sqrt{3}S^4_{\Tilde{x}}-S^4_{\Tilde{y}})$\\
    & $(\sqrt{3}S^1_{\Tilde{x}}+S^1_{\Tilde{y}})-(\sqrt{3}S^2_{\Tilde{x}}+S^2_{\Tilde{y}})+(\sqrt{3}S^3_{\Tilde{x}}+S^3_{\Tilde{y}})-(\sqrt{3}S^4_{\Tilde{x}}+S^4_{\Tilde{y}})$\\
\end{tabular}
}
}
\end{ruledtabular}
\caption{\label{tab:raman-operator} {\bf Raman operators in each symmetry channel.} $x,y,z$ ($\tilde{x},\tilde{y},\tilde{z}$) denote global (local) coordinates axes defined in Fig. 1e.}
\end{table*}

Figure 3b compares the Raman spectra measured at $T$=350 K and $T$=15 K. At $T$=350 K, all six Raman-active phonon modes ($1 A_{1} \oplus 1 E \oplus 4 T_{2}$) are observed, with energies similar to those in Eu$_2$Ir$_2$O$_7$ and Pr$_2$Ir$_2$O$_7$ (Refs.~\citenum{Ued19,Xu22}). Below $T_\textrm{C}$, three additional spin excitation peaks, $A_{1}(R^E)$, $E(R^{A_1})$, and $T_{2}(R^{T_2})$, emerge below the lowest energy phonon mode at 37.6 meV (Fig.~3a). These modes are labeled according to their intrinsic symmetry and the symmetry of the Raman operator in parentheses.  Additionally, Nd crystal electric field (CEF) excitations at $\approx$25 meV (Ref.~\citenum{Xu15}), which persist above $T_\textrm{C}$, become more intense and split below it. These CEF excitations are identified by their lack of a selection rule, appearing in all symmetry channels. The $A_{1}(R^E)$ peak at $\approx$30.1 meV has been observed in Eu$_2$Ir$_2$O$_7$ and assigned to a single-magnon excitation~\cite{Ued19}. This assignment is consistent with the selection rules (Fig.~3a and Table 1), and with the single-magnon mode observed in RIXS (Extended Data Fig.~4). In cubic symmetry, the four single-magnon modes are split into a singlet and a triplet at the $\Gamma$ point~\cite{Lee13}, the latter of which corresponds to the $T_{2}(R^{T_2})$ at $\approx$22.8 meV (Fig.~3a). 

A striking observation is the emergence of an intense $E(R^{A_1})$ mode at $\approx$16.1 meV. Since no linear combination of spin operators transforms as $A_1$, this peak must correspond to a two-magnon excitation (Table 1). The $E(R^{A_1})$ excitation is highly sensitive to impurities and appears as a broad continuum in lower-quality samples (Extended Data Fig.~5). However, the narrow peak width in our high-quality sample indicates formation of a two-magnon bound state, revealing attractive magnon-magnon interactions mediated by quantum fluctuations~\cite{Bet31,Han63, Wor63}. Notably, its energy lies below that of the single magnon mode, a hallmark of unconventional magnetism, suggesting that single magnons are not the fundamental excitations in this system. 

Furthermore, this result suggests that the two-magnon collective mode condenses prior to the onset of AIAO order as the sample is cooled, resulting in a spin nematic phase~\cite{Chu91,Sha06, Zhi10,Pra20, Kim24, She25}. Figure 3c and Extended Data Fig.~6 show the spectra in the $R^{A_1}$ channel below 30 meV where electronic scattering dominates the signal. Upon cooling, this signal increases, peaks at $\sim$205 K, and then decreases, indicating condensation of a collective mode strongly damped by the electron-hole continuum. Around this temperature, splitting of the $E$ phonon mode and the emergence of an additional $T_2$ phonon mode are observed (Fig.~3c), providing clear evidence of cubic symmetry breaking. This contrasts with the AIAO phase of Eu$_2$Ir$_2$O$_7$, where no additonal phonon mode was detected~\cite{Ued19}). 
Thus, our Raman data reveal a nematic order intertwinned with AIAO order below  T$_\textrm{C}$, as shown in Extended Data Fig.~1. 

\vspace{5 mm}
\noindent
\textbf{Mean-field phase diagram}

\begin{figure*}
\includegraphics[width=1\textwidth]{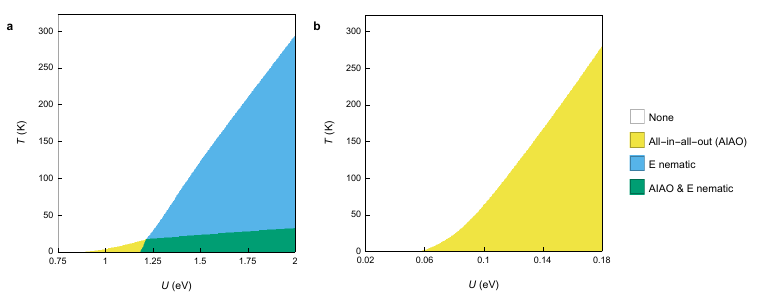}
\caption{\label{fig:wide} {\bf Mean-field phase diagram.} The phase diagrams of a QBT model with phenomenological interactions. {\bf a,} Phase diagram for $J/U=1.018$, $D/U\approx-0.8939$, $V/U\approx-0.7435$, $\Gamma_0/U\approx-0.1173$, and $\Gamma_1/U\approx0.352$. {\bf b,} Phase diagram for $J=D=V=\Gamma_0=\Gamma_1=0$.} 
\end{figure*}


To provide additional theoretical insight into the cubic symmetry breaking orders beyond AIAO, we employ a mean-field approach based on a continuum QBT model.
Previous renormalization group studies have shown that the system exhibits instabilities toward various symmetry breaking orders, including AIAO and nematic orders~\cite{DasSarma2017, Herbut2017, Moessner2021}, suggesting that quadrupolar orders may emerge alongside AIAO in Nd$_2$Ir$_2$O$_7$. To quantitatively compare theory with experiment, we analyze the Landau-Ginzburg free energy expanded in terms of the AIAO and nematic order parameters. The free energy is derived by integrating out electrons in the QBT model with phenomenological interactions, including the Hubbard $U$, Heisenberg exchange $J$, Dzyaloshinskii–Moriya interaction $D$, and symmetry-allowed anisotropic terms $\Gamma_{0,1}$ (Refs.~\citenum{Wit13, Lee13, Brydon2022} and Supplementary note 6). The parameters for the QBT model are similar to those used in previous theoretical studies on pyrochlore iridates~\cite{Wit13}, particularly Nd$_2$Ir$_2$O$_7$ (Ref.~\citenum{Tia16}). The order parameter configurations that minimize the free energy are summarized in Fig.~4. 

At high temperatures, the system retains full cubic and time-reversal symmetries, which are progressively lowered by nematic and AIAO orders. Consistent with previous renormalization group analyses~\cite{DasSarma2017,Moessner2021}, we find that pure AIAO and nematic orders appear adjacent to each other. When the Hubbard interaction $U$ dominates, with $J, D$ of comparable magnitude to $U$, $E$ nematic order emerges first and coexists with AIAO at lower temperatures (Fig.~4a). This stands in stark contrast to the phase diagram of the QBT model with only the Hubbard interaction $U$, where our mean-field calculations find only AIAO (Fig.~4b), consistent with Refs.~\citenum{Wit12, Wit13, Lee13, Berke2018}. The inter-site couplings $J$ and $D$, which enhance inter-site entanglement, promote the formation of nematic order. As the strengths of $J$ and $D$ increase, more complex combinations of $E$ and $T_2$ nematic orders can emerge and  coexist with AIAO (Supplementary Note 6). Regardless of these details, we find a broad range of parameters where nematic orders emerge first as the temperature decreases, followed by AIAO at lower temperatures, in agreement with our Raman data. This provides further theoretical evidence of additional symmetry breaking beyond AIAO.

\vspace{5 mm}
\noindent
\textbf{Discussion}

Our results demonstrate that RIXS interferometry can effectively detect `hidden' orders that remain invisible to most probes (Supplementary Note 7). The spin quadrupoles on the bonds arise from entanglement between nearest-neighbor spins, leaving an observable trace as oscillations in RIXS intensities. Earlier RIXS studies have shown that interference effects arise in solids containing isolated dimers~\cite{Rev19} and tetramers~\cite{Mag24}, where molecular orbitals---a rudimentary form of quantum entanglement confined to small atomic clusters---naturally occur. Our work broadens the applicability of this technique to systems where quantum entanglement exists over an extended lattice, and in phases where quasiparticle description of the electronic structure breaks down~\cite{Nak16}. Although our model is based on a small cluster, which has limited accuracy in predicting excited energy levels, it captures the salient features of interference effects that arise from quantum entanglement when integrated over suitable energy windows.    

Our work thus directly reveals the strong quantum fluctuations, widely believed to mediate Cooper pairing in unconventional superconductors, that underlie the formation of the two-magnon bound state. In cuprates, it is well known that quantum fluctuations reduce the two-magnon excitation energy from 4$J$ to 2.7$J$ (where $J$ is the nearest-neighbor exchange interaction), which is still above the single magnon band~\cite{Blu94}. In our case, the highly unusual nature of the spin state, despite exhibiting long-range order, is manifested as the two-magnon bound state below the single magnon band. This is a rare phenomenon found only in low-dimensional magnets with strong anisotropy~\cite{Tor69}, or in a high-field fully polarized regime~\cite{Fog24}, and it reflects a strong tendency toward pair condensation into a spin nematic phase~\cite{Fog24}. Most recently, a collective excitation associated with a quadrupolar order was observed at the bottom of the single-magnon band in the spin nematic phase of Sr$_2$IrO$_4$ (ref.~\citenum{Kim24}). To the best of our knowledge, a two-magnon bound state has not been observed in a three-dimensional system. 

A key aspect of our findings is that the seemingly distinct broken symmetries do not merely coexist but instead arise from a highly entangled quantum state. Revealing this entangled basis is crucial for understanding both the nature of these orders and the mechanism by which they form. As a result, the standard approach of treating these broken symmetries as separate, competing orders cannot fully capture the richness of the underlying many-body state. Our results have broad implications beyond the scope of pyrochlore iridates, and they complement the growing perspective that multiple phases---including density waves, nematic, time-reversal symmetry-breaking orders and superconductivity---in cuprate and iron-based high-T$_{\textrm C}$ superconductors arise from one underlying principle~\cite{Fra15,Fer19}.

\noindent\textbf{Methods}

\noindent\textbf{Crystal synthesis}
Nd$_2$Ir$_2$O$_7$ single crystals were grown via solid-state epitaxy on yttria-stabilized zirconia (YSZ) seed crystal. A stoichiometric mixture of Nd$_2$O$_3$ and IrO$_2$ powders was thoroughly ground and pelletized. The pellet was placed on a YSZ crystal and vacuum sealed in a fused silica tube. The sealed tube was heated to 1100$^{\circ}$C for 72 hours, followed by furnace cooling to room temperature. After growth, a film-like Nd$_2$Ir$_2$O$_7$ crystal, with a thickness of a few micrometers, was epitaxially formed on the YSZ single crystal. To remove residual powder from the pellet, the crystal was cleaned with ethanol in a sonicator.

\noindent\textbf{Resonant inelastic X-ray scattering}
High-resolution inelastic x-ray scattering experiments were performed at the ID20 beamline of the European Synchrotron Radiation Facility (ESRF) and the 27-ID-B beamline of the Advanced Photon Source (APS). The incident photon energy was tuned to the Ir $L_3$ pre-edge (11.215 keV). At ESRF, a Si (1\,1\,1\,) high-heat-load monochromator in combination with a Si~(8\,4\,4) channel-cut monochromator reduced the energy bandpass down to 14.6\,meV. Similarly, at APS, a diamond (1\,1\,1\,) high-heat-load monochromator was used with the same channel-cut monochromator setup, achieving an equivalent energy bandpass. In both cases, the beam was focused using Kirkpatrick-Baez mirrors, producing a spot size of approximately $10\times10$ (H$\times$V)\,\textmu m$^2$ FWHM at the sample position. Scattered photons were analyzed by a Si~(8\,4\,4) diced spherical analyzer with a radius of 2\,m. A 60/,mm mask was used in front of the diced analyzer for data collected at ESRF (Fig. 2a, e-g, and 6\,K data in Fig. 2h), while a 40\,mm mask is used for the 50\,K data in Fig. 2h at APS. The overall energy resolution was about 30\,meV. A horizontal scattering geometry was used with the incident $\pi$-polarization and the outgoing polarization was not resolved. To reduce the elastic tail from the Bragg peaks at (0 0 2$n$), measurements were conducted along the (0.3, 0.3, $L$).

\noindent\textbf{RIXS calculation}
Many-body calculations were performed using a three-band ($t_{2g}$-band) Hubbard model on a tetrahedron cluster. We consider the microscopic Hamiltonian incorporating the spin-orbit coupling($\lambda_{SOC}$), trigonal distortion($\Delta_{tr}$), Kanamori-type Coulomb interactions($U$ and $J_H$), and hopping integrals between nearest neighboring Ir's ($t_{\sigma}, t_{\pi}, t_{\delta}, t_{oxy}$). 
The parameters are optimized to fit the experimental RIXS spectra assuming $t_{\pi} = -2t_{\sigma}/3 = 0.2$, $t_{oxy}=0.4$ and $t_\delta$ = 0, with $U =2.0, J_H=0.4, \lambda_{SOC}=0.5, \Delta_{tr}=-0.5$, in units of eV. The optimized hopping parameters satisfy the AIAO ground conditions such that $-1.2 < t_{\sigma} / t_{oxy} = -0.75 < -0.5$ (Ref.~\citenum{Lee13,Wit13}).
By solving the Hamiltonian using the exact Lanczos diagonalization method, we obtain the doublet ground states, which correspond to the AI and AO states. We calculate the RIXS spectra by employing the Kramers-Heisenberg formula under fast collision and dipole approximations. In these approximations, the RIXS intensity is calculated by the dynamical correlation functions of the RIXS operator (see Supplementary Note 4 for details). 


\noindent\textbf{Raman spectroscopy}
Raman spectroscopy was performed on a home-built instrument equipped with a 750-mm spectrometer and a liquid nitrogen-cooled CCD detector using a 633-nm laser as excitation source. Bragg grating notch filters were used to suppress elastic signals, allowing measurement of inelastic signals above 5 cm$^{-1}$ with a resolution of 0.33 cm$^{-1}$. Parallel and crossed polarizations of incident and scattered lights propagating along the $c$-axis were measured in a backscattering geometry. 
The laser power and beam-spot size are kept below 0.25 mW and 2 \textmu m, respectively. 
All Raman spectra are corrected for the Bose factor.

\noindent\textbf{Resonant x-ray diffraction}
Resonant x-ray diffraction (RXD) experiments were conducted at the 1C beamline of the Pohang Light Source-\MakeUppercase{\romannumeral 2}~\cite{Kim23}. The incident photon energy was set to 11.215 keV, corresponding to the Ir $L_3$ pre-edge, to probe the magnetic Bragg peak at (0, 0, 10). The beam was focused to $25 \times 20$ \textmu m$^2$ using Kirkpatrick–Baez mirrors. The sample was cooled down to 5 K using a closed-cycle cryostat.
 
\noindent\textbf{Mean-field calculation}
The phase diagram is determined by minimizing the Landau-Ginzburg free energy for a QBT model with phenomenological interactions. The QBT model is given by 
\begin{align*}
    H = \alpha\beta k^2\mathds{1} - 2\alpha\cos{\chi}\left(\frac{\sqrt{3}(k_x^2-k_y^2)}{2}\gamma_1+\frac{3k_z^2-k^2}{2}\gamma_2\right)\\
    - 2\alpha\sin{\chi}\left(\sqrt{3}k_yk_z\gamma_3+\sqrt{3}k_zk_y\gamma_4+\sqrt{3}k_xk_y\gamma_5\right).
\end{align*}
Here, $\{\gamma_a\}_{a=1,\cdots,5}$ are second-rank $J=3/2$ spin tensors~\cite{Moessner2021} and $(\alpha,\beta, \chi)$ are the parameters determined by projecting the tight-binding Hamiltonian near the $\Gamma$ point~\cite{Wit13, DasSarma2017, Brydon2022}. The tight-binding parameters are $t_{oxy} \approx 0.8852\text{ eV}$, $t_\sigma \approx -1.010 \text{ eV}$, $t_\pi \approx 0.7014\text{ eV}$ and $t_{\sigma(\pi)}'/t_{\sigma(\pi)} \approx 0.04326$, which is similar to those in Refs.~\citenum{Wit13, Tia16}. Phenomenological interactions are introduced into this model by projecting the Hubbard term $U$, the nearest-neighbor Heisenberg interaction $J$, the nearest-neighbor Dzyaloshinskii–Moriya interaction $D$, the nearest-neighbor charge-charge interaction $V$, and the nearest-neighbor anisotropic interactions $\Gamma_{0,1}$~\cite{DasSarma2017, Brydon2022, Wit13, Lee13}. The Fierz identity~\cite{Herbut2017} is then applied to decompose the resulting four-fermion interaction terms into AIAO and nematic order parameters. Next, fermionic degrees of freedom are integrated out up to one-loop order and to quartic terms in the order parameters. The phase diagram is obtained by minimizing the resulting free energy.


\noindent\textbf{Data availability}

All data are available in the main text or the supplementary information.

\noindent\textbf{Acknowledgements}
We thank Kyusung Hwang, E.~G. Moon, Natasha Perkins, Yuan Li, and Igor Herbut for insightful discussions. This work was supported by the National Research Foundation (NRF) of Korea grant funded by the Korea government (RS-2024-00360303), Basic Science Research Program through the NRF of Korea funded by the Ministry of Education (2022R1I1A1A01056493) and Samsung Science and Technology Foundation under Project Number SSTF-BA2201-04. The use of the Advanced Photon Source at the Argonne National Laboratory was supported by the U. S. DOE under Contract No. DE-AC02-06CH11357. We acknowledge the European Synchrotron Radiation Facility (ESRF) for provision of synchrotron radiation facilities under proposal number HC-4054 and we would like to thank F. Gerbon for assistance and support in using beamline ID20. S.J. and G.Y.C. are supported by Samsung Science and Technology Foundation under Project Number SSTF-BA2002-05 and SSTF-BA2401-03, the NRF of Korea (Grants No.~RS-2023-00208291, No.~2023M3K5A1094810, No.~2023M3K5A1094813, No.~RS-2024-00410027, No.~RS-2024-00444725) funded by the Korean Government (MSIT), the Air Force Office of Scientific Research under Award No.~FA2386-22-1-4061, and Institute of Basic Science under project code IBS-R014-D1. BHK was supported by the Institute for Basic Science in the Republic of Korea under Project No. IBS-R024-D1. The works at Seoul National University were supported by the Leading Researcher Program of the National Research Foundation of Korea (Grant No. RS-2020-NR049405).

\noindent\textbf{Author contributions}
B.J.K. conceived and managed the project. B.H.K and G.Y.C planned the theoretical project. Juny.K. grew single crystals and performed RIXS experiments with help from C.J.S., A.L. and Jung.K.; J.K. and B.J.K performed symmetry analysis; G.O., K.K. and H.K. performed Raman experiments; Juny.K., J.K., S.H., H.-W.J.K. performed RXD experiments; S.J. and G.Y.C. performed mean-field calculations with help from G.S. and A.G.; B.H.K performed exact diagonalization calculations; B.W., F.A., G.G., N.M., M.M. and M. L. T. performed XRD experiments. B.J.K. wrote the manuscript with inputs from all authors.

\noindent\textbf{Competing interests}

Authors declare that they have no competing interests.

\setcounter{figure}{0}
\makeatletter 
\renewcommand{\figurename}{\textbf{Extended Data Fig.}}
\renewcommand{\tablename}{\textbf{Extended Data Table}}
\makeatother

\begin{figure*}
\includegraphics[width=180 mm]{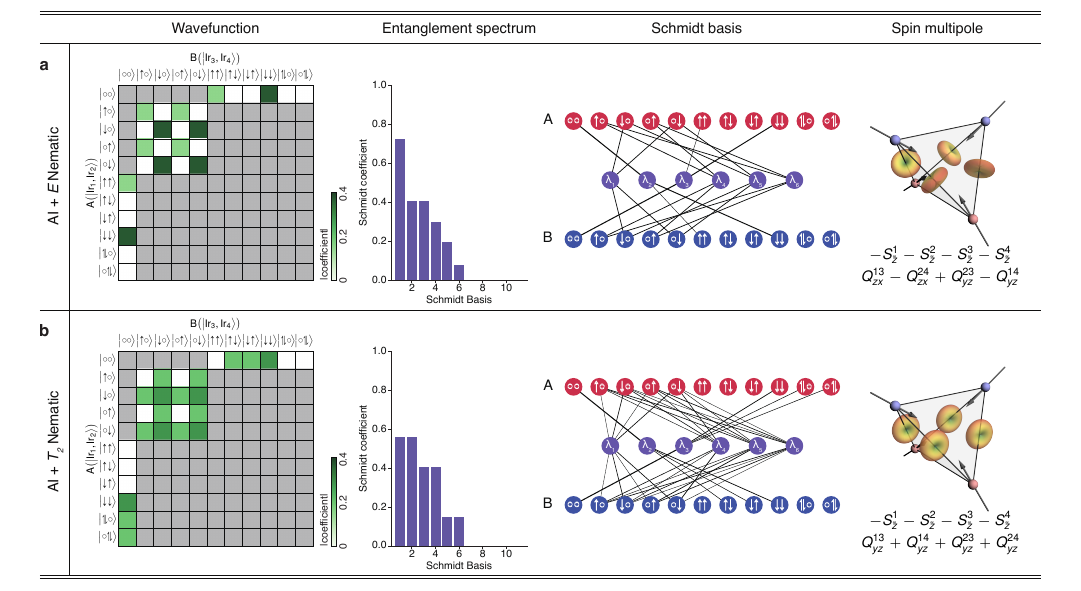}

\caption{\label{fig:wide} {\bf AI+$\boldsymbol{E}$ nematic and AI+$\boldsymbol{T_2}$ nematic orders and their entanglement structures.} Wavefunction, Schmidt coefficients, Schmidt basis states, and spin multipoles for states with AI+$E$ nematic ({\bf a}), and AI+$T_2$ nematic ({\bf b}) orders. 
}
\end{figure*}

\begin{figure*}
\includegraphics[width=180 mm]{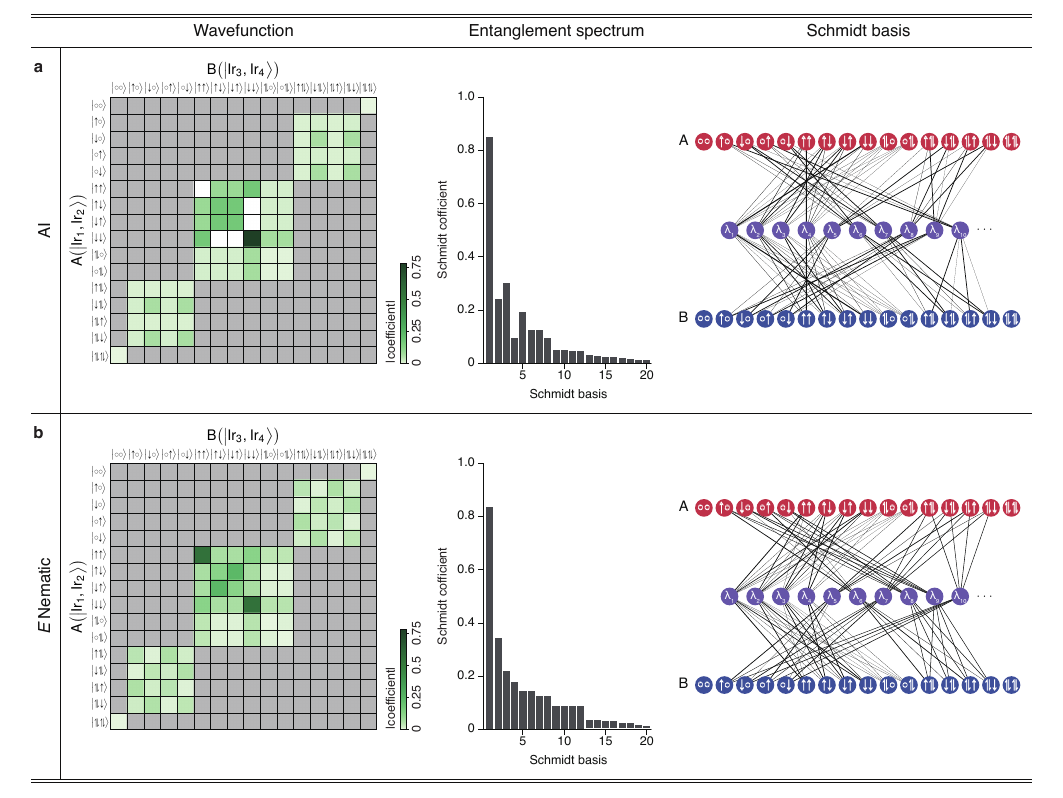}

\caption{\label{fig:wide} {\bf AI+$\boldsymbol{E}$ nematic and AI+$\boldsymbol{T_2}$ nematic orders and their entanglement structures.} Wavefunction, Schmidt coefficients, Schmidt basis states, and spin multipoles for states with AI+$E$ nematic ({\bf a}), and AI+$T_2$ nematic ({\bf b}) orders. Given the large number of contributing states, only the 54 basis states with the largest weights are shown.
}
\end{figure*}

\begin{figure*}
\includegraphics[width=175 mm]{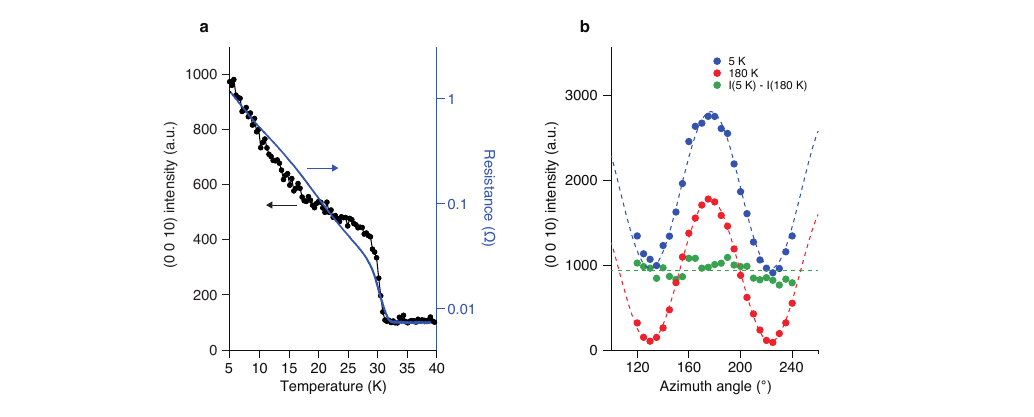}

\caption{\label{fig:wide} {\bf AIAO order measured in the sample} {\bf a,} The temperature dependence of (0 0 10) magnetic Bragg peak measured (blue) at an azimuthal angle of 135$^{\circ}$ (defined to be zero when (1 0 0) is in the scattering plane), where anisotropic tensor of susceptibility (ATS) scattering is minimized. Corresponding phase transition observed by four-probe resistance measurements (black) are overlaid forcomparison. {\bf b,} Azimuthal-angle dependence of the (0 0 10) magnetic Bragg peak intensity. The contribution from the AIAO order to the scattering intensity is isolated by taking the difference (green) between the profiles measured above (red) and below (blue) T$_\textrm{C}$. The AIAO order exhibits no azimuthal angle dependence\cite{Don16}.
}
\end{figure*}

\begin{figure*}
\includegraphics[width=175 mm]{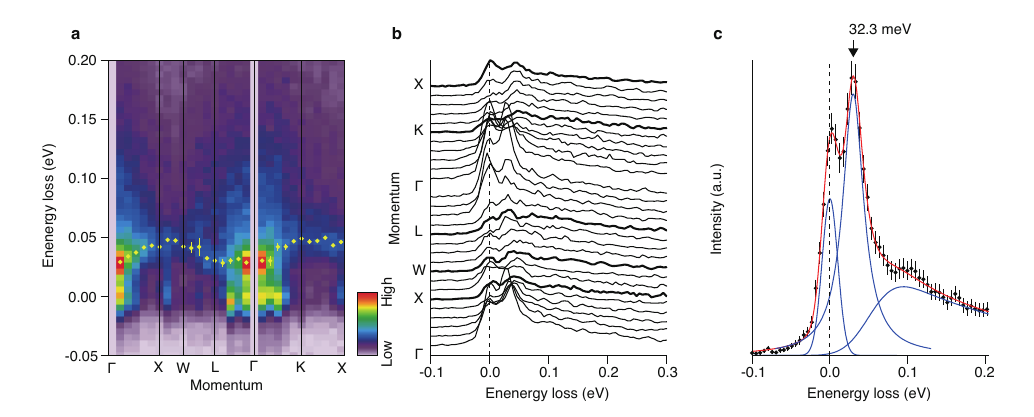}

\caption{\label{fig:wide} {\bf Spin wave dispersion measured by RIXS.} {\bf a,} False color map of RIXS intensity measured at 6 K along high symmetry points. Yellow circles indicate single-magnon energies determined from fitting the data. {\bf b,} Stack plot of the energy spectra. The spectra are measured around $\Gamma$ point at (2 -2 12).  {\bf c,} $\Gamma$ point spectrum is fitted to elastic (Gaussian), magnon (Lorentzian), and higher-energy excitations (Gaussian convoluted with exponential decay).
}
\end{figure*}

\begin{figure*}
\includegraphics[width=84mm]{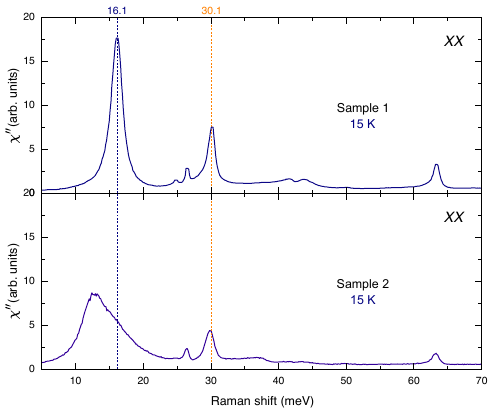}

\caption{\label{fig:wide} {\bf Sample dependence of $E(R^{A_1})$ mode.} Comparison of the spectra taken from samples grown at the optimal growth condition (Sample 1) and at a suboptimal condition (Sample 2). 
}
\end{figure*}

\begin{figure*}
\includegraphics[width=53mm]{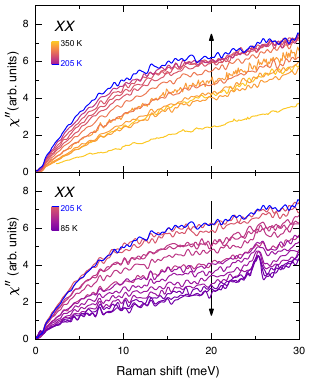}
\caption{\label{fig:wide} {\bf Temperature dependence of quasi-elastic scattering.} Raman spectra below 30 meV measured at temperatures above(upper) and below(lower) 205 K. The total intensity peaks at 205 (blue). CEF of Nd at $\approx$25 meV is resolved at low temperatures.
}
\end{figure*}


\end{document}